\documentclass[%
 aip,
 amsmath,amssymb,
 reprint,%
]{revtex4-2}

\usepackage{graphicx}
\usepackage{dcolumn}
\usepackage{bm}

\usepackage[utf8]{inputenc}
\usepackage[T1]{fontenc}
\usepackage{mathptmx}

\begin{document}


\title{Collective steady-state patterns of swarmalators with finite-cutoff interaction distance}

\author{Hyun Keun \surname{Lee}}
\affiliation{Department of Physics, Sungkyunkwan University, Suwon 16419, Korea}
\author{Kangmo \surname{Yeo}}%
\affiliation{Department of Physics, Jeonbuk National University, Jeonju 54896, Korea}
\author{Hyunsuk \surname{Hong}}
\email{hhong@jbnu.ac.kr}
\affiliation{Department of Physics, Jeonbuk National University, Jeonju 54896, Korea}
\affiliation{Research Institute of Physics and Chemistry, Jeonbuk National University, Jeonju 54896, Korea}

\date{\today}

\begin{abstract}
We study the steady-state patterns of population of the coupled oscillators that sync and swarm, where the interaction distances among oscillators have finite-cutoff in interaction distance. We examine how the static patterns known in the infinite-cutoff are reproduced or deformed, and explore a new static pattern that does not appear until a finite-cutoff is considered. All steady-state patterns of the infinite-cutoff, static sync, static async, and static phase wave are respectively repeated in space for proper finite-cutoff ranges. Their deformation in shape and density takes place for the other finite-cutoff ranges. Bar-like phase wave states are observed, which has not been the case for the infinite-cutoff. All the patterns are investigated via numerical and theoretical analysis. 
\end{abstract}

\maketitle

\begin{quotation}
Swarming~\cite{swarm} is the collective behavior of biological/artificial entities in the absence of centralized coordination. Interaction between agents is the reason for the behavior, and thus its modeling is a core task to understand or control the emergent phenomena. Recently, a steady state pattern was studied~\cite{Fetecau11} for the overdamped limit model with potential while motivated by biological aggregations in viscous liquid, and the static feature later becomes various~\cite{Kevin17} as self-propelling factor is implemented through the phase dynamics of synchronization~\cite{Pikovsky03}. The interaction therein is mediated by a potential function that may change in time. We here consider the cutoff of interaction distance to model the fact that the distance any living/engineering entities can manage is necessarily finite. The previously understood static feature lasts till the cutoff is larger than the size of the pattern formed in the absence of cutoff. For the smaller cutoff, the pattern is deformed in an explainable manner, and then it becomes amorphous or non-stationary for further decrease. A new class of static structures not appearing in the no-cutoff model is observed and specified. Phase diagram for the steady-state patterns according to the cutoff distance is illustrated. Finite-cutoff of interaction distance makes the swarm behavior prolific as combined with underlying potential.
\end{quotation}

\section{\label{int}introduction}

Synchronization in the population of coupled oscillators has been widely explored.  In particular, not only mathematical models but also various experimental systems have been considered to investigate the synchronization behavior~\cite{Winfree67,Kuramoto84,Kuramoto87,Crawford94,Crawford95,Crawford99, Strogatz03, Pikovsky03, Acebron05, Hong05,Hong07, HS11}. In one recent study~\cite{Kevin17}, the oscillators that also swarm~\cite{Fetecau11,swm-bio,swm0,swm-atf2,swm-sp,swm-spp1d,swm,swm-bio3,swm-bio2,swm-atf,swm2} was introduced while called {\textit{swarmalator}}, and the self-organization in space and time was reported. Three steady states and two nonstationary ones have been found in the swarmalators model with long-range attraction and repulsion. The swarmalator model is followed by various researches~\cite{Kevin18-pre,Kevin19-cp,Ha19-MMM,Bettstetter19-IEEE,Bettstetter20-IEEE,Smith20-pre,Lizarraga20-chaos,Morales20-pre} in its novel modeling of collective behaviors and practical applicability as well.

According to Refs.~\cite{Fetecau11,Kevin17}, the spatial sizes of the steady-state patterns are finite when the interaction distance is long-ranged of no cutoff or infinite-cutoff. Here, one may expect there can be a finite-cutoff in interaction range, for which the result is qualitatively same in the sense that all the steady-state patterns by the long range interaction are reproduced. If a specific pattern known for the infinite-cutoff is the target of biological or engineering swarms, a proper interaction-range cutoff larger than the pattern's size could be enough. If an interaction range is smaller than this cutoff but still comparable, a describable change of the pattern is expected. Moreover, since the realization of infinite range interaction is by itself impractical, the system with finite-cutoff has actual implications. The various spatiotemoral constraints~\cite{Bettstetter19-IEEE,Bettstetter20-IEEE} are indeed the factual barriers in realizing the notion of swarmalators. All of these motivate the current work.

In this paper, we consider the local interaction among the swarmalators, where the interaction range is restricted by a control parameter, and explore the collective steady states of the system. We examine how the steady-state patterns of the system with long-range interaction are deformed in the understandable way and also pay special attention to the possibility of new steady states. This paper consists of five sections.  Section~\ref{model} introduces the model, and Sec.~\ref{ssic} revisits the system with infinite-cutoff (global interaction).  In Sec.~\ref{fci}, the finite-cutoff  interaction is considered, and the various steady-state patterns are demonstrated. Section~\ref{sum} gives a brief summary.

\section{\label{model}model}

\begin{figure}
\includegraphics[width=\columnwidth]{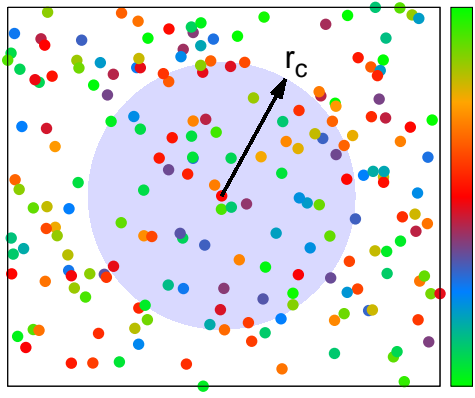}
\caption{(Color Online) Schematic diagram for the one swarmalator's neighbors controlled by the range parameter $r_c$. Here the dots represent the swarmalators and the colors their phases $\theta (\in [-\pi, \pi))$.}
\label{fig:interaction_range}
\end{figure}

We consider a population of oscillators governed by
\begin{eqnarray}
\dot{\bf r}_i
&=& \frac{1}{N_i(r_{\rm c})}\sum_{j \in \Lambda_i(r_{\rm c})}
\left[\Big(1+J\cos(\theta_{j}-\theta_{i})\Big) - \frac{1}{r^2_{ij}}\right]
\nonumber \\
&&~~~~~~~~~~~~~~~~~~~ \times\Big({\bf r}_j-{\bf r}_i\Big),
\label{eq:model_x} \\
\dot{\theta}_{i}
&=& \omega_i +
\frac{K}{N}\sum_{j \neq i}
\frac{\sin(\theta_{j}-\theta_{i})}{r_{ij}},
\label{eq:model_theta}
\end{eqnarray}
where ${\bf r}_i = {\bf r}_i(t) =(x_i(t), y_i(t))$ is the position vector of agent $i$ at time $t$ in 2-dimensional space with $r_{ij} \equiv |{\bf r}_i-{\bf r}_j|$ and $\theta_i=\theta_i(t)$ is the $i$'s phase angle. Note that the dynamics governed by Eqs.~(\ref{eq:model_x}) and (\ref{eq:model_theta}) makes the oscillators swarm and sync.  Considering such characteristic we here call the oscillators {\it{swarmalators}}, following the Ref.~\cite{Kevin17}. $N_i(r_{\rm c})$ is the number of members in the collection $\Lambda_i(r_{\rm c})$ composed of such swarmalators whose distances to $i$ are not greater than $r_c$ (see Fig.~\ref{fig:interaction_range}). This way, the interaction range of spatial dynamics is bounded by $r_{\rm c}$.
$J$ is the parameter for attraction with $|J|\le 1$, for which $J=1$ is used in the numerical work below.
$\omega_i$ is the natural frequency of $i$, $K$ is a coupling strength of phase and $N$ is the total number of swarmalators in system. We consider identical swarmalators with $\omega_i=0$ for all $i$ in this work.

In the $r_{\rm c}\to\infty$ limit, when $J=0$, Eq.~\eqref{eq:model_x} recovers the model suggested in Ref.~\cite{Fetecau11} of special interest in the formation of uniform density and identical radius disc of swarms. The motion of ${\bf r}_i$ for $J=0$ is basically a gradient flow as each summand can be rewritten by the gradient of a scalar function for $r_{ij}$. As Eq.~\eqref{eq:model_x} with $J=0$ is motivated by the biological aggregations in viscous liquid~\cite{swm,swm2}, the inertia effect is ignored to set up an overdamped limit equation~\cite{Strogatz1994}. Next when $J\neq 0$, Eqs.~\eqref{eq:model_x} and \eqref{eq:model_theta} corresponds to one of the models in Ref.~\cite{Kevin17}, which share common feature of various active and static phases of the swarmalators. Due to the phase variables in Eq.~\eqref{eq:model_x}, which change in time by Eq.~\eqref{eq:model_theta}, the motion of ${\bf r}_i$ now has self-propelling property~\cite{swm0,swm-atf2,swm-sp}. Equation~\eqref{eq:model_theta} is a variant of the Kuramoto model~\cite{Kuramoto84,Kuramoto87}, i.e., the coupling strength $K$ in this prototype model is generalized to $K_{ij}=K/r_{ij}$ in Eq.~\eqref{eq:model_theta} in order to reflect the spatial configuration in phase dynamics. This was suggested in Ref.~\cite{Kevin17} to generate pairing current of the swarmalators in a proper range of negative $K$ with $r_{\rm c}=\infty$.

\section{\label{ssic}steady states for infinite cutoff}

According to Refs.~\cite{Fetecau11,Kevin17}, there are three steady-state patterns when $r_{\rm c}=\infty$. As we are interested in their reproduction or moderate change for finite $r_{\rm c}$, we begin with explaining the three patterns, in a comprehensive way as possible. In the steady states, $\dot r_i=0$ and $\dot \theta_i=0$ for all $i$. Therefore, as its continuum counterpart such as velocity field ${\bf v}({\bf r})$ at position ${\bf r}$, it is reasonable to set ${\bf v}({\bf r})=0$. When one considers a proper coarse-grain of Eq.~\eqref{eq:model_x} in the steady states, the velocity field reads
\begin{equation}
\label{cv}
{\bf v}({\bf r})=\int d{\bf r}'\left(-\nabla_{{\bf r}} U({\bf r},{\bf r}') \right)\rho({\bf r}')=0,
\end{equation}
where
\begin{eqnarray}
\label{KR}
U({\bf r},{\bf r}')&=&
\frac{1}{2}|{\bf r} - {\bf r}'|^2 (1+J\cos(\theta({\bf r})-\theta({\bf r}'))) -\ln|{\bf r} - {\bf r}'| \nonumber \\
&\equiv&
U_{\rm att}({\bf r},{\bf r}') + U_{\rm rep}({\bf r},{\bf r}')
\end{eqnarray}
with phase field $\theta({\bf r})$, and $\rho({\bf r})$ is the steady-state density of swarmalators at position ${\bf r}$ with normalization $\int d{\bf r}\rho({\bf r})=1$. As Eq.~\eqref{cv} holds for all ${\bf r}$, it also holds that $\nabla_{\bf r}\cdot {\bf v}({\bf r})=0$. Then, in the fact that $\nabla_{\bf r}^2 U_{\rm rep}({\bf r},{\bf r}')=2\pi \delta({\bf r}-{\bf r}')$~\cite{Arfken}, one finds a self-consistent equation for $\rho({\bf r})$, which is given by
\begin{eqnarray}
\label{keq}
\rho({\bf r})&=& \frac{1}{2\pi}\int d{\bf r}'\nabla_{\bf r}^2  U_{\rm att}({\bf r},{\bf r}') \rho({\bf r}').
\end{eqnarray}
This self-consistent equation plays a central role in finding the steady-state patterns in the help of other information like symmetry and/or numerical observations.

One of the three steady-state patterns appears when $K>0$. For positive $K$, Eq.~\eqref{eq:model_theta} derives the system to phase synchronization state of a common phase $\theta_{\rm s}$. Then, Eq.~\eqref{eq:model_x} becomes free from phase as $\cos(\theta_j-\theta_i)\to1$ as
evolution goes on. In the steady state, therefore, it reads
\begin{equation}
\label{Uas}
U_{\rm att}({\bf r},{\bf r}')=\frac{1}{2}|{\bf r}-{\bf r}'|^2(1+J).
\end{equation}
When this is considered in Eq.~\eqref{keq} in the two-dimensional $(x, y)$ coordinate, since $\nabla_{\bf r}^2 U_{\rm att}({\bf r},{\bf r}')=2(1+J)$ is constant, using $\int d{\bf r}'\rho({\bf r}')=1$, $\rho({\bf r}) = (1+J)/\pi$ is immediate. Next, it should be answered where to reside.

\begin{figure}
\includegraphics[width=\columnwidth]{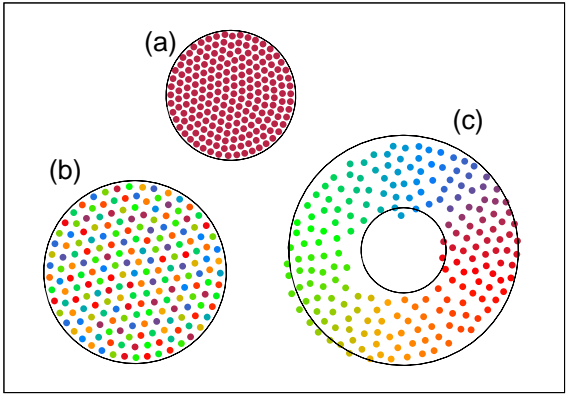}
\caption{(Color Online) Three steady-state patterns known in the $r_{\rm c}\to\infty$ limit~\cite{Kevin17}: (a) is the static sync (SS) when $K>0$. (b) is the static async (SA) when $K<K_{\rm c}$ for a negative critical value $K_{\rm c}$. (c) is the static phase wave (SPW) when $K=0$. For numerical realization, $N=200$ system with $J=1$ is considered. $K=1$ and $K=-2$ are used for (a) and (b), respectively. The black circles therein guide the associated radius analytically understood (see text).}
\label{nss}
\end{figure}

The center of positions is conserved by the symmetric pairs of changes in Eq. (1). Thus  when the center is used as the origin, it is natural to expect the rotational symmetry of density, i.e., $\rho({\bf r})= \rho(r)$ for $r=|{\bf r}|$. The numerical results always give such distributions that strongly suggest a circular $\rho({\bf r})$ as shown in Fig.~\ref{nss}(a).
Thus when this is considered in Eq.~\eqref{keq}, it follows that
\begin{equation}
\label{ds}
\rho({\bf r}) = \frac{1}{\pi R_{\rm s}^2}~~{\rm for}~~ r < R_{\rm s},
\end{equation}
where $R_{\rm s}$ is given by
\begin{equation}
\label{Rs}
R_{\rm s}=1/\sqrt{1+J}.
\end{equation}
Interestingly, the density is uniform and the radius depends only on $J$. The pattern by Eqs.~\eqref{ds} and \eqref{Rs} was named the static sync (SS)~\cite{Kevin17}.

The steady-state condition [Eq.~\eqref{cv}] holds for Eq.~\eqref{ds} as follows. When
\begin{equation}
\label{V}
V({\bf r})=\int d{\bf r}' U({\bf r},{\bf r}') \rho({\bf r}')
\end{equation}
is introduced, $\nabla_{\bf r}\cdot {\bf v}({\bf r}) =0$ corresponds to Laplace's equation $\nabla_{\bf r}^2V({\bf r}) =0$. Thus, $\rho({\bf r})$ of Eq.~\eqref{keq} gives $V({\bf r})$ of the Laplace's equation, and this is mediated by Eq.~\eqref{V}. As $V({\bf r})=V(r)$ by $U_{\rm att}({\bf r},{\bf r}')=U_{\rm att}(|{\bf r}-{\bf r}'|)$ in Eq.~\eqref{Uas} and $\rho({\bf r})=\rho(r)$ in Eq.~\eqref{ds}, the solution of the Laplace's equation is given by such cylindrical harmonics of no longitudinal neither angular dependence. Then, the choice should be~\cite{Arfken} $V({\bf r})\propto 1$ exclusively or $\ln r$. Here, one excludes the latter as $V({\bf r})$ remains finite at ${\bf r}=0$ in a simple calculation of Eq.~\eqref{V} with Eqs.~\eqref{Uas} and \eqref{ds}. Thus, ${\bf v}({\bf r})=-\nabla_{\bf r}\cdot V({\bf r})=0$ is immediate for constant $V({\bf r})$.

The second steady-state pattern appears~\cite{Kevin17} when $K<K_{\rm c}<0$ holds for a negative critical value $K_{\rm c}\approx -1.2J$. For negative $K$, uniform $\theta_i$ is unstable in Eq.~\eqref{eq:model_theta}, and an erratic profile (asynchrony) would appear. Below $K_{\rm c}$, if the configuration of $\theta_i$ is erratic enough, the summation of the sinusoidal part of Eq.~\eqref{eq:model_x} may vanish. When this is the case, $U_{\rm att}({\bf r},{\bf r}')=(1/2)|{\bf r}-{\bf r}'|^2$ reads. This is no more than Eq.~\eqref{Uas} with $J=0$, from which it follows that
\begin{equation}
\label{Ra}
\rho({\bf r}) = \frac{1}{\pi}
~~{\rm for}~~r<R_{\rm a}=1.
\end{equation}
This result of uniform density and unit radius successively explains the numerical data shown in Fig.~\ref{nss}(b). This pattern was named the static async (SA)~\cite{Kevin17}.

The third steady-state pattern appears~\cite{Kevin17} when $K=0$. For $K=0$, one trivially reads $\dot\theta_i = 0$ in Eq.~\eqref{eq:model_theta}. The numerical data shown in Fig.~\ref{nss}(c) indicates $\theta_i$ can be replaced with the spatial angle $\phi_i$ of ${\bf r}_i$; $\theta_i = \phi_i + c$ for a constant $c$. Then, in polar coordinate ${\bf r} = (r,\phi)$, it reads
\begin{eqnarray}
\label{Upw}
U_{\rm att}({\bf r},{\bf r}')
&=& \frac{1}{2}\left(r^2+r'^2-2rr'\cos(\phi-\phi')\right) \nonumber \\
&& ~~ \times \left(1+J\cos(\phi-\phi')\right).
\end{eqnarray}

Using Eq.~\eqref{Upw}, we find its Laplacian is given by
\begin{eqnarray}
\nabla_{\bf r}^2 U_{\rm att}({\bf r},{\bf r}')  &=& 2-\frac{Jr'}{2r} +
f_1(r,r')\cos(\phi-\phi') \nonumber\\
&& ~~~~~~ + f_2(r,r')\cos2(\phi-\phi')
\label{eq:LapUatt}
\end{eqnarray}
for proper functions $f_1(r,r')$ and $f_2(r,r')$. Also, the numerically observed circular strip suggests that $\rho({\bf r})=\rho(r)$. Plugging this and Eq.~\eqref{eq:LapUatt} into Eq.~\eqref{keq}, one finds
\begin{equation}
\label{ssp3}
\rho({\bf r}) = \frac{1}{\pi} - \frac{C}{r}~~{\rm for}~~R_1<r<R_2
\end{equation}
for constant $C$. One again considers Eq.~\eqref{ssp3} in Eq.~\eqref{keq} to know
\begin{equation}
\label{C1}
C=\frac{2J(R_2^3-R_1^3)}{3\pi(4+J(R_2^2-R_1^2))}.
\end{equation}
Normalization of $\rho({\bf r})$ gives
\begin{equation}
\label{C1n}
(2\pi C - R_1-R_2)(R_1-R_2)=1.
\end{equation}

To fix the solution, one may refer to Eq.~\eqref{cv} of ${\bf v}({\bf r})=0$. This requires vanishing radial component because ${\bf v}({\bf r}) = (v_{\rm att}(r)\ + v_{\rm rep}(r))\hat {\bf r}$ in $U_{\rm att/rep}({\bf r},{\bf r}')=U_{\rm att/rep}(|{\bf r}-{\bf r}'|)$ and $\rho({\bf r})=\rho(r)$, where $\hat {\bf r}\equiv {\bf r}/r$ is the unit vector in radial direction. Attraction part is
\begin{eqnarray}
\label{vatt}
v_{\rm att}(r)&=&\int d{\bf r}' (-\partial_r U_{\rm att}({\bf r},{\bf r}'))\rho(r')
\\
&=& (R_2^2-R_1^2-2\pi C (R_2-R_1))r - 2\pi C. \nonumber
\end{eqnarray}
For repulsion part, one can use the divergence theorem~\cite{Arfken} and $\nabla_{{\bf r}''}^2 U_{\rm rep}({\bf r}'',{\bf r}')=2\pi \delta({\bf r}''-{\bf r}')$ to write
\begin{eqnarray}
\label{vrep}
2\pi r v_{\rm rep}(r)
&=& -\int_{R_1 \le r''\le r}
d{\bf r}'' \int d{\bf r}' \nabla_{{\bf r}''}^2 U_{\rm rep}({\bf r}'',{\bf r}')\rho(r') \nonumber \\
&=& 2\pi \left(2\pi C(r-R_1)-(r^2-R_1^2)\right).
\end{eqnarray}

With Eqs.~\eqref{vatt} and \eqref{vrep}, one knows $v_{\rm att}(r)+v_{\rm rep}(r)=0$ is possible only when
\begin{equation}
\label{cv0}
2\pi C = R_1.
\end{equation}
Using Eq.~\eqref{cv0}, one obtains in Eqs.~\eqref{C1} and \eqref{C1n} that
\begin{eqnarray}
\label{R1R2}
R_1 &=& \Lambda_J((12-J)\sqrt{3}-3\sqrt{(4+J)(12-5J)})/12J \nonumber \\
R_2 &=& \Lambda_J/2\sqrt{3}
\end{eqnarray}
for $\Lambda_J \equiv ((12-3J+\sqrt{3(4+J)(12-5J)})/(2-J))^{1/2}$. Then, Eqs.~\eqref{cv0} and \eqref{R1R2} fixes the solution in Eq.~\eqref{ssp3}. This was named the static phase wave (SPW)~\cite{Kevin17}.

\section{\label{fci}Finite-cutoff of interaction distance}
In the numerical study with finite $r_{\rm c}$, we have observed various steady-state patterns. Some of them are same as those in the $r_{\rm c}\to\infty$ limit, another ones are their proper deformations, and the others are classified as a new class. The three classes appear in order as $r_{\rm c}$ decreases, and the shape becomes amorphous or non-stationary for further decrease. In the following, the observed patterns are understood and/or described by exploiting Eq.~\eqref{keq}, the self-consistent equation for density.

\subsection{\label{sd}SS for $r_{\rm c}>2 R_{\rm s}$}
\begin{figure}
\includegraphics[width=\columnwidth]{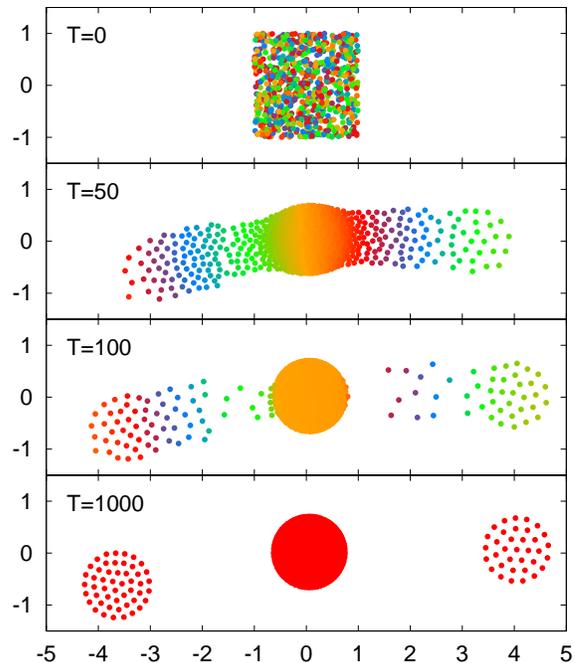}
\caption{(Color Online) Formation of sync state pattern for $r_{\rm c}=1.5$ in $N=800$ system with $K=0.1$ and $J=1$ in the long time limit. Each panel is a snapshot at the specified time $T$ at left upper corner. Multiple SS discs are finally formed as shown in the last panel. For initial condition, uniformly random positions sampled out of $L\times L$ square at origin was used for $L=2$ (this type of initialization is applied below in this work).}
\label{rc1.5kp}
\end{figure}

For finite $r_{\rm c}>2 R_{\rm s}$, the system with positive coupling strength $K$ is found to be in SS. It is interesting to note that the steady state with multiple discs in SS is discovered (see Fig.~\ref{rc1.5kp}).  Multiple discs of SS have not been found in the system with $r_{\rm c}=\infty$. Figure~\ref{rc1.5kp} is the numerical results of $N=800$ system for $r_{\rm c}=1.5$ when $K=0.1$ and $J=1$ (we below fix $J=1$), displaying the process how the multiple discs of SS are formed.
The radius of the discs is, in common, $\approx1/\sqrt{2}$ that reminds us of $R_{\rm s}$ in Eq.~\eqref{Rs} with $J=1$. A simple explanation of this numerical observation is following; the swarmalators are divided into three groups apart from others farther than $r_{\rm c}$ during transient period, and each group evolves into its own steady state.  The snapshots are shown in Fig.~\ref{rc1.5kp}, from initial $T=0$ via transient $T=50,~100$ to final $T=1000$.

Once a number of swarms is included in a circular region of radius $r_{\rm c}$ and this region is separated from the others farther than $r_{\rm c}$, the dynamics taking place thereafter in the region is identical to that by $r_{\rm c}=\infty$. If there are several regions separated that way (the number of regions depends on initial condition), the same number of patterns finally appears. We note, when $r_{\rm c}>2R_{\rm s}$, the self-consistent equation for density [Eq.~\eqref{keq}]  does not change from that for $r_{\rm c}=\infty$. Then, each final pattern becomes no more than the SS of Eqs.~\eqref{ds} and \eqref{Rs}.  As expected, the discs in the last snapshot are, in common, of radius $1/\sqrt{2}$ with their own uniform densities. The minimal separation between the discs is $2R_{\rm s}$ as a consequence of the condition $r_{\rm c}>2R_{\rm s}$.

With no loss of generality, the condition $r_{\rm c}>2R_{\rm s}$ considered for SS case is generalized to
\begin{equation}
\label{c1}
r_{\rm c}>D_\infty~,
\end{equation}
where  $D_\infty$ is the diameter of a pattern appearing in the $r_{\rm c}\to\infty$ limit. It can be SA or SPW beside SS depending on $K$. That is, any of them is expected in the same reason when Eq.~\eqref{c1} holds and, when repeated, the minimal separation is given by the diameter.

\subsection{\label{ad}SA for $r_{\rm c}>2 R_{\rm a}$}
\begin{figure}
\includegraphics[width=\columnwidth]{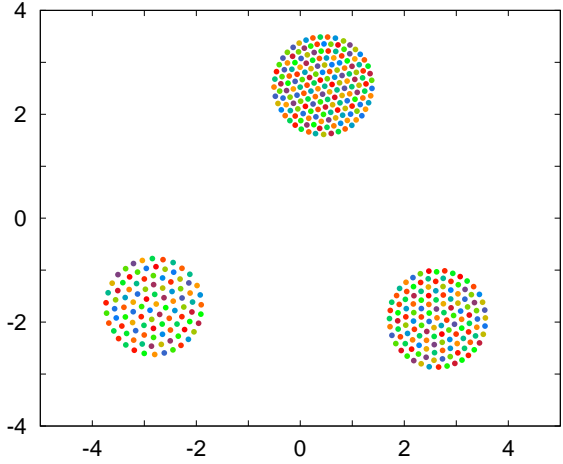}
\caption{(Color Online) Multiple SA discs in $N=400$ system with $r_{\rm c}=2.1$ when $K=-8$ and $J=1$. For initial condition, that for Fig.~\ref{rc1.5kp} was used with $L=10$.}
\label{fig:async}
\end{figure}

In case of static async, $D_\infty=2 R_{\rm a}$. Thus by Eq.~\eqref{c1}, SA should appear for finite $r_{\rm c}>2 R_{\rm a}$ when the negatively strong coupling $(K<K_c<0)$ is provided. As expected, the system is found to be in SA where the phases of all the swarmalators are random in the full range of $[-\pi, \pi )$. Also, similar to the previous sync case, the steady state composed of multiple discs of SA is discovered, which has not been found in the system with $r_{\rm{c}}=\infty$. Figure~\ref{fig:async} shows multiple SA discs for $r_{\rm c}=2.1$ (note $D_\infty=2R_{\rm a}=2$ for SA).

\subsection{\label{asd}Anomalous SS/SA with nonuniform radial density}
\begin{figure}
\includegraphics[width=\columnwidth]{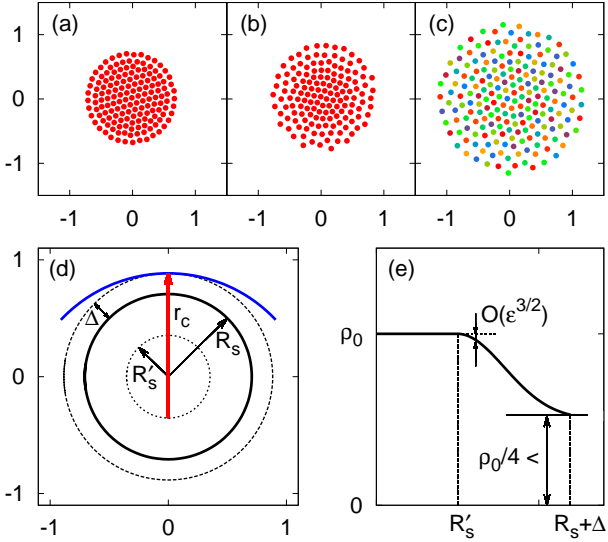}
\caption{(Color Online) Anomalous static states of $N=200$ system and supporting schematics: Anomalous static sync (aSS) in (a) and (b) are, respectively, the case when $r_{\rm c}=1.3$ and $r_{\rm c}=1.2$ for $K=0.1$ and $J=1$. Anomalous static async (aSA) in (c) is when $r_{\rm c}=1.7$ for $K=-2$ and $J=1$. The geometric condition required for these patterns are depicted in (d). The red arrow therein guides to the eyes, showing the relation given by Eq.~(\ref{drprc}). Schematic of radial density is illustrated in (e). In (d) and (e), $R_{\rm s}$ and $R_{\rm s}'$ are replaced with $R_{\rm a}$ and $R_{\rm a}'$, respectively, in the async case.}
\label{rc1.25kp}
\end{figure}

Next, we find there appear anomalous steady-state patterns for $r_{\rm c}<2 R_{\rm s}$ when $K>0$, as shown in Fig.~\ref{rc1.25kp}(a) and (b).
Because Eq.~\eqref{c1} does not hold this time for $D_\infty=2 R_{\rm s}$, what a different pattern may form was in question, and the shown anomalous disc is the result to be understood. The radius becomes larger to $R_{\rm s}+\Delta$. The density looks uniform inside the region of radius $R_{\rm s}'(< R_{\rm s})$ while decreases thereafter. These observations can be explained with Eq.~\eqref{keq} as follows.

For finite $r_{\rm c}$, it is given from Eq.~\eqref{Uas} that
\begin{equation}
\label{Urc}
\nabla_{\bf r}^2 U_{\rm att}({\bf r},{\bf r}')=2(1+J)H(r_{\rm c}-|{\bf r}-{\bf r}'|),
\end{equation}
where $H(x)$ is the Heaviside step-function assigned with $1$ for $x \ge 0$, or $0$ otherwise. Plugging Eq.~\eqref{Urc} into Eq.~\eqref{keq} with the numerical observation $\rho({\bf r})=\rho(r)$, one may consider the geometric situations by $R_{\rm s}+\Delta$, $R_{\rm s}'$,  and $r_{\rm c}$. Figure~\ref{rc1.25kp}(d) shows a situation when the interaction range of radius $r_{\rm c}$ marginally covers the whole distribution of radius $R_{\rm s}+\Delta$; see the blue arc and the dashed circle tangentially meet. Then, since $\nabla_{\bf r}^2 U_{\rm att}({\bf r},{\bf r}')$ is constant in the interaction range, Eq.~\eqref{keq} gives $\rho(r=R_{\rm s}')=(2\pi)^{-1}2(1+J)\int d{\bf r}'H(r_{\rm c}-|{\bf r}-{\bf r}'|)\rho(r') =((1+J)/\pi)\int d{\bf r}'\rho(r') = (1+J)/\pi$. This value does not change as long as the blue circle covers the dashed one, and this is the case when $r\le R_{\rm s}'$. Note $(1+J)/\pi$ is the very $(\pi R_{\rm s}^2)^{-1}$ of Eq.~\eqref{ds}.

When $r > R_{\rm s}'$, the blue circle cannot include the dashed one, and the covered region decreases as $r$ increases. So that, in Eq.~\eqref{keq}, the integration range becomes smaller for $r$, and this leads to decreasing $\rho(r)$ in the left hand side. Then, one may write
\begin{equation}
\label{drp}
\rho({\bf r})=
\begin{cases}
\rho_0
&{\rm for}~r\le R_{\rm s}' \\
g(r)&{\rm for}~R_{\rm s}'<r<R_{\rm s}+\Delta~,
\end{cases}
\end{equation}
where $\rho_0\equiv(\pi R_{\rm s}^2)^{-1}$ and $g(r)$ is a proper function that decreases for $r$ from $g(R_{\rm s}')=\rho_0$. Figure~\ref{rc1.25kp}(e) shows the schematic of Eq.~\eqref{drp}. The density begins to decrease from $\rho_0$ following $O(\epsilon^{3/2})$ for small $\epsilon=r-R_{\rm s}'>0$ as the area  $\sim \epsilon^{3/2}$ is excluded from the integration in Eq.~\eqref{keq}. A lower bound of $g(R_{\rm s}+\Delta)$ is simply arguable in the $R_{\rm s}'=0^+$ case, which is accompanied with $r_{\rm c}=R_{\rm s}+\Delta$. In this case, when the blue circle is moved to locate its center at $r=R_{\rm s}+\Delta$, Eq.~\eqref{keq} gives $g(R_{\rm s}+\Delta)> \rho_0/4$ as the integration range covers, at least, a quarter sector of the disc in the dashed circle. Note the coverage is minimized when $R_{\rm s}'=0^+$, so that $\rho_0/4$ can be a lower bound of $g(R_{\rm s}+\Delta)$. As the truncated $U_{\rm att}({\bf r},{\bf r}')$ by $r_{\rm c}$ still shows $U_{\rm att}({\bf r},{\bf r}')=U_{\rm att}(|{\bf r}-{\bf r}'|)$ and  the density in Eqs.~\eqref{Urc} and \eqref{drp} also does $\rho({\bf r})=\rho(r)$, $|V(0)|<\infty$ follows. Then, a consequent constant $V({\bf r})$ gives  ${\bf v}({\bf r})=-\nabla_{\bf r}V({\bf r})=0$. Hence, the density in Eq.~\eqref{drp} can be a steady-state configuration. We call this anomalous static sync (aSS).

The aSS explained so far is conditioned in $r_{\rm c}<2 R_{\rm s}$. Meanwhile, the numerical data manifest there should be a lower bound on  $r_{\rm c}$. For a survey on lower bound, we revisit Fig.~\ref{rc1.25kp}(d). It states
\begin{equation}
\label{drprc}
R_{\rm s}+R_{\rm s}'+\Delta=r_{\rm c}
\end{equation}
with $R_{\rm s}'=R_{\rm s}'(r_{\rm c})$ and $\Delta=\Delta(r_{\rm c})$. As $R_{\rm s}$ is independent of $r_{\rm c}$, the lower bound of $r_{\rm c}$ is reduced to that of $R_{\rm s}'(r_{\rm c})+\Delta(r_{\rm c})$. Also in the independence, Eq.~\eqref{drprc} shows $R_{\rm s}'(r_{\rm c})+\Delta(r_{\rm c})$ is decreasing for $r_{\rm c}$. Since $R_{\rm s}'$ approaches $0$ as $r_{\rm c}$ decreases, the lower bound $l$ of $r_{\rm c}$ is $l=R_{\rm s}+\Delta(l)$. Thus when $r_{\rm c}=l$, it reads $\rho(r)=g(r)$ in the absence of the central constant region. For a simple estimation of $l$, we regard $g(r)$ is linear and $g(l)\approx \rho_0/4$. From normalization $\int d{\bf r}\rho(r)=1$, it follows $l\approx \sqrt{2}R_{\rm s}$. Thus considering the above-mentioned upper bound, one finds an interval
\begin{equation}
\label{itvad}
\sqrt{2} R_{\rm s} \lesssim r_{\rm c} < 2 R_{\rm s}
\end{equation}
for the formation of aSS, and this is, interestingly, consistent with our numerical data. In the other side in $r_{\rm c} \lesssim \sqrt{2} R_{\rm s}=1$ (note we set $J=1$ in numerics), we numerically observe the pattern becomes amorphous. $l \approx \sqrt{2}R_{\rm s}$ is by itself the upper bound of $R_{\rm s}+\Delta$, the radius of aSS. Depending on initial condition, there can appear multiple aSS. After transient period, each region separated farther than $r_{\rm c}$ from the others has its own disc (not shown here).

Our argument so far is also applicable for the change of the asynchronized-swam disc for $K<K_{\rm c}<0$ while $R_{\rm s}$ is replaced with $R_{\rm a}$. We call the pattern anomalous static async (aSA). An instance is shown in Fig.~\ref{rc1.25kp}(c). We remark the async-counterpart of Eq.~\eqref{itvad}, $\sqrt{2} R_{\rm a} \lesssim r_{\rm c} < 2 R_{\rm a}$ is valid for the smaller criterion than $K_{\rm c}$. In the numerical data obtained for $N=400$, the lower bound shows deviation till $K\gtrsim -2$ (see the upper-left part of the aSA region in Fig.~\ref{diag}). There is more deviation for larger $K$ and smaller $r_{\rm c}$ in the corner. This observation is consistent with the reported stable-region of SA in Ref.~\cite{Kevin17} for $r_{\rm c}=\infty$, $K\lesssim K_{\rm c}\approx -1.2J$, in that the criterion $K_{\rm c}$ is a consequence of linear stability and a smaller $r_{\rm c}$ means a stronger perturbation. Different from aSS case, aSA does not become amorphous but a non-stationary pattern when it disappears by the decrease of $r_{\rm c}$.

\subsection{\label{vrd}Variations in SPW}
\begin{figure}
\includegraphics[width=\columnwidth]{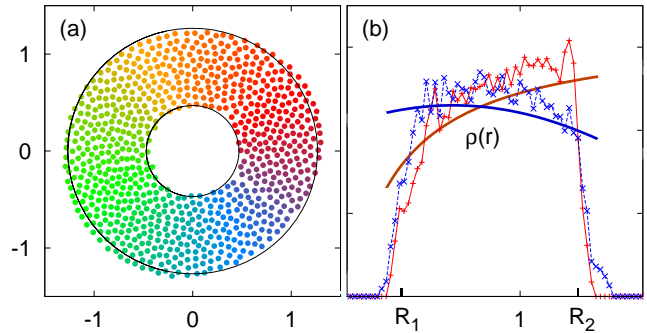}
\caption{(Color Online) Variations of SPW by finite $r_{\rm c}$ in $N=800$ system with $K=0$ and $J=1$: (a) shows the formed SPW for $r_{\rm c}=1.6$ slightly exceeds the region between the radius $R_1$ and $R_2$ of Eq.~\eqref{R1R2} known for $r_{\rm c}=\infty$. In (b), radial densities for $r_{\rm c}=2.5$ (red) and $r_{\rm c}=1.6$ (blue) are shown. The red pluses and blue crosses represent the numerical data, and the bold curves show the overall profile in each cases.}
\label{rclessR2}
\end{figure}

The diameter of SPW that appears when $K=0$ for $r_{\rm c}=\infty$ is $D_\infty=2R_2$. Thus by Eq.~\eqref{c1}, for $r_{\rm c}>2R_2$, a formation of the same SPW(s) is expected, and this is numerically observed. Next, for $r_{\rm c}<2R_2$, the results become different, also as expected. However, this time, the difference is not that apparent in shape; the inner and outer radii change a little though the cutoff value is decreased a lot to $r_{\rm c}\approx 1.6$ from $2.5 \approx 2R_2$. Figure~\ref{rclessR2}(a) is a pattern at $r_{\rm c}=1.6$, whose inner and outer radii are not discriminated so clearly from $R_1$ and $R_2$, respectively, known for $r_{\rm c}=\infty$ in Eq.~\eqref{R1R2}.

Instead, we have found that a substantial change in density takes place while $r_{\rm c}$ decreases a lot. In Fig.~\ref{rclessR2}(b), the red pluses come from the numerical data obtained at $r_{\rm c}=2.5$ that is slightly smaller than $2 R_2= 2.53..\,$. Each of them is the count of swarms in a narrow circular region between radii $r-\delta$ and $r+\delta$, where $r$ is discrete and small $\delta>0$ is a few times of the resolution of $r$. Their overall profile is guided by the bold red curve from Eqs.~\eqref{ssp3}, \eqref{cv0}, and \eqref{R1R2}. This guidance is plausible as $r_{\rm c}=2.5 \approx 2 R_2$. The blue crosses in Fig.~\ref{rclessR2}(b) come from the numerical data at $r_{\rm c}=1.6$, and the blue bold curve is their overall shape as a guide to eyes. The substantial change in density is supposed to compensate for the small change in shape despite the considerable decrease of $r_{\rm c}$. We also observed the small change of shape mainly occurs below $r_{\rm c}\approx 1.8$. The inner and outer radii remain apparently same for $1.8\lesssim r_{\rm c} < 2 R_2$ while the density change takes place.

A heuristic argument for this observation is following. Let $r_{\rm c}$ be smaller that $2R_2$, and consider a circle of radius $r_{\rm c}$ whose center is in somewhere in SPW. If the center is near the outer boundary of SPW, the circle cannot cover the SPW. When this incomplete coverage is considered in Eq.~\eqref{keq}, the reduced integral region would lead to  density decrease at the center. Next, consider the center is near the inner boundary. In this case, the $r_{\rm c}$-radius circle covers the SPW when $r_{\rm c}-R_2>R_1$ is provided. For the density of such center, the integral region of Eq.~\eqref{keq} is conserved, and there is no explicit reason to decrease the density. Instead, an opposite reason was already due because the expected density decrease near the outer region requires mass conservation in the system.
We consider the cross between the density profiles shown in Fig.~\ref{rclessR2}(b) is the consequence.

If $r_{\rm c}-R_2>R_1$ does not hold, the integral region of Eq.~\eqref{keq} is always reduced whatever points of SPW is used for the center of $r_{\rm c}$-radius circle. This may indicate a termination of the shape-preserving density change. One then considers an interval
\begin{equation}
\label{itvr}
R_1+R_2 \lesssim r_{\rm c} < 2 R_2\,,
\end{equation}
where such SPWs of negligible/considerable change in radius/density for $r_{\rm c}$ are respectively expected. The lower bound suggested this way is rather comparable with the numerical lower bound $r_{\rm c}\approx 1.8$ in that $R_1+R_2 =  1.74.. \approx 1.8$ (see Eq.~\eqref{R1R2} with $J=1$ for the $R_1+R_2$ value). Thus when SPW is specified by the shape only, the condition for its formation is relaxed from $r_{\rm c}>2 R_2$ to $r_{\rm c}\gtrsim R_1 + R_2$.

We remark that some cases do not give SPW when $r_{\rm c}$ begins to be smaller than $1.8$, depending on initial conditions. No appearance of SPW below $r_{\rm c}\approx 1.8$ becomes more frequent for the smaller $r_{\rm c}$. Furthermore, SPW does not appear when $r_{\rm c} \lesssim 1.5$, at least, numerically. In brief, the numerical data suggests that the basin of attractor~\cite{Strogatz1994} for SPW begins to decrease from $r_{\rm c}\approx 1.8$, and then disappears below $r_{\rm c}\approx 1.5$.

\subsection{\label{sbar}Bar-like patterns at $K=0$}
\begin{figure}
\includegraphics[width=\columnwidth]{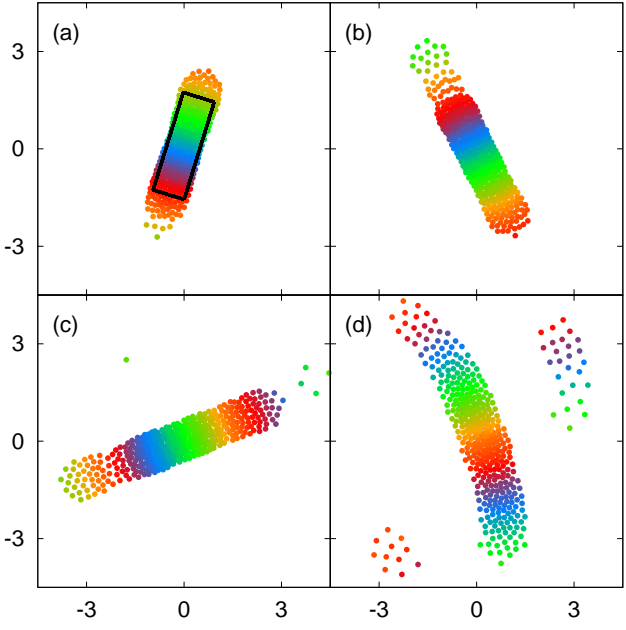}
\caption{(Color Online) Various bar-like patterns (Bar) for $1.2\lesssim r_{\rm c} \lesssim 1.8$ in $N=400$ system with $K=0$ and $J=1$: (a)-(d) are obtained for $r_{\rm c}=1.8, 1.6,1.4$, and $1.2$, respectively. The black rectangle in the pattern of (a) is of length $\pi$ and width $1$.}
\label{fbar}
\end{figure}

It is worthy of noting that, when the pattern's shape is not circular below $r_{\rm c}\approx 1.8$, it looks linear. According to the numerical data, linear shapes begin to appear for $r_{\rm c}\approx 1.8$, and disappear when $r_{\rm c}\lesssim 1$. Figure~\ref{fbar} shows a few linear bar-like patterns (Bar).
It is inappropriate we consider to view such patterns as the deformed APW by finite $r_{\rm c}$. As a first step to the understanding of such bar-like patterns, we below argue that each of them could be a deformed shape, by its own finite $r_{\rm c}$, of the bar given as the solution of Eq.~\eqref{keq} in the $r_{\rm c}\to\infty$ limit. An ironical situation here is no bar-like pattern has been observed yet in our numerical study neither reported in literature as far as we know. We thus conjecture that a solution bar exists for Eq.~\eqref{keq}, of no stability, but it becomes stable for finite $1\lesssim r_{\rm c}\lesssim 1.8$ accompanied with proper deformation. If our scenario is valid, the pattern of Fig.~\ref{fbar}(a) is most close to the solution bar among the figures therein because its $r_{\rm c}=1.8$ is larger than the others'.

We choose $x$-axis as the direction of bar with no loss of generality. In the numerical data, phase smoothly increases along the $x$-axis. The positive direction of the $x$-axis is selected in line with the phase increase. We thus introduce phase field $\theta({\bf r})=\theta(x)$ that is differentiable for $x$, and then write
\begin{eqnarray}
\label{Ubar}
U_{\rm att}({\bf r},{\bf r}')&=&-\frac{1}{2}\left((x-x')^2+(y-y')^2 \right) \nonumber \\
&& ~~~~ \times \left(1+J\cos(\theta(x)-\theta(x'))\right).
\end{eqnarray}
in rectangular coordinate. No dependence on $r_{\rm c}$ shows Eq.~\eqref{Ubar} is that for the infinite cutoff.

Consider a bar of length $2l$ and of width $2w$, whose center is the origin. By the symmetry of $\rho(x,y)=\rho(|x|,|y|)$, we choose $\theta(0)=0$ for simplicity. We then give a relation between $\rho(x,y)$ and $\theta(x)$ as
\begin{equation}
\label{rt}
\int dy\,\rho({\bf r}) = \rho(x)=\frac{1}{2\pi}\frac{d\theta}{dx}
\end{equation}
in the consideration of i) $\theta_i$s are initialized with uniform random variables in our numerics and ii) any of them does not change in time in Eq.~\eqref{eq:model_theta} with $K=0$. By Eq.~\eqref{rt}, $\theta(x)$ increases from $\theta(0)=0$ to $\theta(l)=\pi$, and then remains same thereafter, i.e., $\theta(x)=\pi$ for $x>l$. By differentiabilty, $d\theta(l)/dx=0$ is given. By the symmetry mentioned above, it reads that $\theta(-x)=-\theta(x)$.

Plugging Eqs.~\eqref{Ubar} and \eqref{rt} into Eq.~\eqref{keq}, after a little algebra, one can obtain
\begin{equation}
\label{rb}
\rho(x,y)=\frac{1}{\pi}-\frac{J}{2\pi^2}\frac{d^2}{dx^2}\Big(c(l)x\sin\theta+s(l)\cos\theta\Big)
\end{equation}
for $c(l)=l+\int_0^l dx\cos\theta$ and $s(l)=\int_0^l dx x\sin\theta$. Obviously, $\rho(x,y)$ of Eq.~\eqref{rb} is considered for $|x|<l$ and $|y|<w$ while  $\rho(x,y)=0$ elsewhere.

One integrates Eq.~\eqref{rb} to know the area $S$ of the solution bar is
\begin{equation}
\label{area}
S=4wl = \pi.
\end{equation}
Interestingly, the pattern in Fig.~\ref{fbar}(a) covers a bar of area $\pi$ by length $\pi$ and width $1$, with a rather tolerable margin (see the guiding black rectangle). This numerical observation suggests $w\approx 1/2$ and $l\approx \pi/2$. We also observed the pattern becomes elongated and broadening as $r_{\rm c}$ decreases. Thus the patterns in Fig.~\ref{fbar}(b)-(d) for the smaller $r_{\rm c}$ also cover the rectangle with the more margins (one can imagine the rectangle in each ones without difficulty). This is consistent with our scenario for the birth of bar-like patterns with finite $r_{\rm c}$, proposed above. We also observe in our numerical data that the bar-like pattern is so deformed for $r_{\rm c}\lesssim 1$ that it becomes amorphous.

The rectangle we have specified is a consequence of numerically motivated speculation encapsulated in Eq.~\eqref{rb}. This equation is again converted with Eqs.~\eqref{rt} and \eqref{area} to
\begin{equation}
\label{de-theta}
\frac{d\theta}{dx}=\frac{\frac{2\pi}{J}\left(x-l\theta/\pi
\right)-c(l)\sin\theta}{c(l)x\cos\theta-s(l)\sin\theta}\,.
\end{equation}
Its solution will lead to the self-consistent equations for $s(l)$ and $c(l)$, which can fix the value of $l$ and next that of $w$ in use of Eq.~\eqref{area}. This way, finding the size of the bar is dual to knowing its density profile. The shooting method looks one of the practical approaches for finding $l$.
We leave related interesting task for future work as remarking $l\approx \pi/2$ and $w\approx 1/2$ is expected with the numerical data up to now.

We add that an 1-dimension result could be relevant to the size of the bar. In 1-dimension, $U_{\rm rep}(r,r')=-|r-r'|$ giving $\nabla_r^2 U_{\rm rep}(r,r') = 2\delta(r-r')$ is responsible for the repulsion part. Thus when $U_{\rm att}(r,r')=(1/2)(r-r')^2(1+J)$ is considered, uniform distribution of width $W(J)=2/(1+J)$ follows. We here consider the swarmalators on $y$-direction segment in the bar are effectively in 1-dimension in that their phases are same so that $U_{\rm att}({\bf r},{\bf r'})=(1/2)(y-y')^2(1+J)$, the distribution is uniform in the segment as written in Eq.~\eqref{rb}, and its width in numerics with $J=1$ is $2w \approx 1=W(1)$. We leave expected explicit connection between the 1-dimension result and the bar for future work.

The steady-state condition Eq.~\eqref{cv} is a sufficient condition for Eq.~\eqref{keq}. That is, the solution of Eq.~\eqref{keq} is not necessarily a steady state. We note $\rho(x)$ of Eq.~\eqref{rb} is not a steady state. For this, one can check the $y$-component of $v_{\rm att}({\bf r})$ given by $\int d{\bf r}' (-\partial_y U_{\rm att}({\bf r},{\bf r}')) \rho({\bf r}')=-y$ cannot balance with its repulsion-counterpart for such $\rho(x)$ that vanishes at $x=l$ (note $\rho(l)\propto d\theta(l)/dx=0$). However, with $r_{\rm c} = 1.8$, we observe a steady-state pattern that is rather close to the rectangle expected with Eq.~\eqref{rb}, as depicted in Fig.~\ref{fbar}(a). We interpret that the bar we speculated with $r_{\rm c}=\infty$ is properly deformed for $1\lesssim r_{\rm c} \lesssim 1.8$, and then it becomes stable while still keeping linear shape more or less. The patterns in Fig.~\ref{fbar} are the instances.

\section{\label{sum}Summary}
\begin{figure}
\includegraphics[width=\columnwidth]{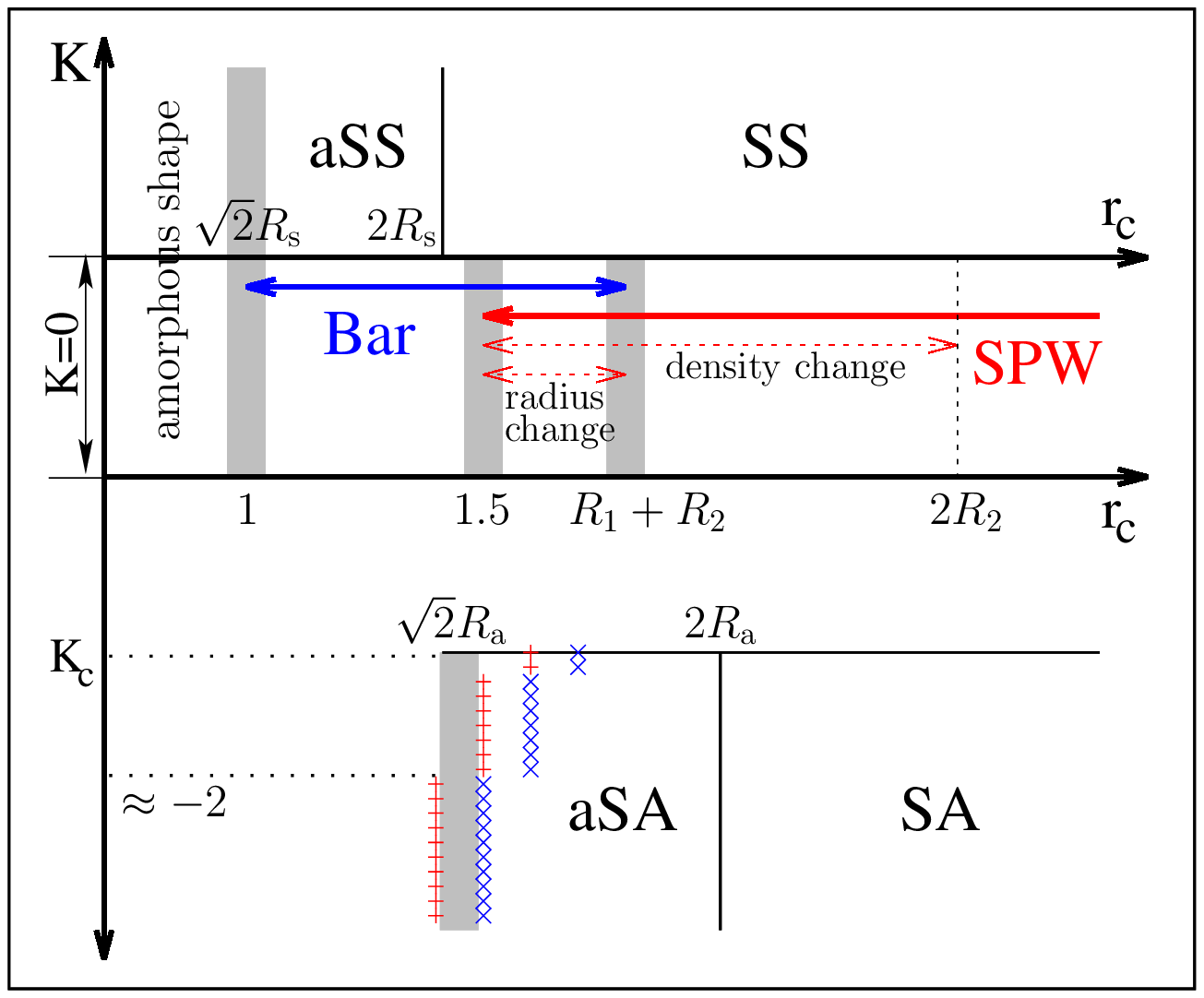}
\caption{(Color Online) Diagram for steady-state patterns for $J=1$ in $(K,r_{\rm c})$ plane: Boundaries are drawn with solid lines while approximated ones with grey blocks. Diagram for $K=0$ is depicted in the $K=0$ band. The boundary by $K_c$ is the consequence of the stability analysis in Ref.~\cite{Kevin17}.
The numerical values for the lower bounds of Bar/SPW and that below $K_c$ are specific to $J=1$ only. The data points around the left-upper part of aSA show the numerical boundary of aSA near there (aSA patterns appear at blue crosses while not at red pluses for $N=400$). The other data points below indicate the argued region of aSA is numerically realized for $K\lesssim -2$.}
\label{diag}
\end{figure}

In this paper, we considered the population of coupled oscillators that sync and swarm under finite-ranged interaction. We explored how the interaction cutoff affects the steady states of the system, with particular attention to the patterns of the steady states. We found that new patterns are induced by the finite-ranged interaction. When the positive/negative phase coupling is present (in the negative case, the coupling below a critical value is considered), multiple clusters forming uniform sync/async discs are found in the static sync/async states. Anomalous discs with nonuniform radial density were also found. In the absence of the phase coupling, APW with abnormal radial density and bar-like patterns of the phase wave state are discovered. We analyzed the new patterns theoretically and numerically to understand their birth and character. The results are summarized in Fig.~\ref{diag}.

The patterns found for $K\ge 0$ are static in the fact the dynamics becomes a gradient flow in the long time limit. But, it is not trivial whether all the patterns for $K< 0$ are stationary because the phase dynamics is frustrated for negative $K$. In this situation, we gave an argument for aSA based on the stability of SA in $K\lesssim -1.2 J$, reported in Ref.~\cite{Kevin17}. The lower bound of $r_{\rm c}$ for aSA, assessed by $\sqrt{2}R_{\rm a}$ in our argument, does not reach $\sqrt{2}R_{\rm a}$ when $K\gtrsim -2$, as shown in Fig.~\ref{diag}. The numerical data obtained at the red-cross points near there but still in $r_{\rm c}\gtrsim \sqrt{2}R_{\rm a}$ give completely different patterns that hardly look like instances of aSA. Those are rather comparable with bar-like ones (not shown here). This implies that that part does not belong to aSA but to non-stationary region, on which understanding is out of the scope of this work. A systematic investigation on the left and upper boundary of aSA could be a starting point in studying the non-stationary states by finite $r_{\rm c}$. We believe that the steady-state features reported in this work will provide their own starting points in extending the understanding to non-stationary region. Related works will appear elsewhere.

\begin{acknowledgments}
We thank to Agata Barcis and Prof. Christian Bettstetter for useful
discussions in the early stage of the study.
This study was supported by NRF-2018R1A2B6001790 and 2021R1A2B5B01001951 (H.H), and 2018R1D1A1B07049254 (H.K.L).
\end{acknowledgments}

\

\noindent{DATA AVAILABILITY}

The data that support the findings of this study are available
within the article.

\

\noindent{REFERNCES}

\nocite{*}
\bibliography{refc}

\end{document}